\documentclass{aa}
\usepackage{amssymb,amsbsy,amsmath}
\usepackage{verbatim,placeins}
\usepackage{graphicx}
\usepackage[varg]{txfonts}
\usepackage[normalem]{ulem} 
\usepackage{aas_macros}

\begin{document}

\title{SPICE PSF Correction: General Framework and Capability Demonstration}
\titlerunning{SPICE PSF Correction Preprint}
\author{J.~E.~Plowman\inst{\ref{inst1}}\and F.~Auch\`{e}re\inst{\ref{inst2}}\and R. Aznar Cuadrado\inst{\ref{inst3}}\and A. Fludra\inst{\ref{inst4}} \and T. Fredvik\inst{\ref{inst5}} \and D. M. Hassler\inst{\ref{inst1}} \and S. Mandal\inst{\ref{inst3}} \and H. Peter\inst{\ref{inst3}}
}
\institute{Southwest Research Institute, Boulder, CO 80302, USA\email{jplowman@boulder.swri.edu}\label{inst1}
	\and
	Université Paris-Saclay, CNRS, Institut d'Astrophysique Spatiale, F-91405, Orsay, France\label{inst2}
	\and
	Max-Planck-Institut für Sonnensystemforschung, Göttingen, Germany\label{inst3}
	\and
	RAL Space, UKRI STFC Rutherford Appleton Laboratory, Didcot, Oxfordshire, OX11 0QX, UK\label{inst4}
	\and
	Institute of Theoretical Astrophysics, University of Oslo, Oslo, Norway\label{inst5}
}
\date{arXiv Preprint}
       
\abstract{We present a new method of removing PSF artifacts and improving the resolution of multidimensional data sources including imagers and spectrographs. Rather than deconvolution, which is translationally invariant, this method is based on sparse matrix solvers. This allows it to be applied to spatially varying PSFs and also to combining observations from instruments with radically different spatial, spectral, or thermal response functions (e.g., SDO/AIA and RHESSI). It was developed to correct PSF artifacts in Solar Orbiter SPICE, so the motivation, presentation of the method, and the results revolve around that application. However, it can also be used as a more robust (e.g., WRT a varying PSFs) alternative to deconvolution of 2D image data and similar problems, and is relevant to more general linear inversion problems.}

\keywords{< line: profiles - techniques: imaging spectroscopy - instrumentation: spectrographs - techniques: high angular resolution - Sun: abundances - Sun: corona >}

\maketitle
	\section{Introduction}

			SPICE \citep[Spectral Imaging of the Coronal Environment,][]{SPICEInstrument_AA2020} is an EUV slit scanning spectrograph on board the Solar Orbiter spacecraft \citep{SolarOrbiter_AA2021}. It has a $\sim 1.1$ arcsecond per pixel spatial plate scale along the slit and several slit widths ranging from 2 to 30 arcseconds, while the spectral plate scale is $\sim 0.009$ nm per pixel. The focal plane has two detector arrays of $1024\times 1024$ pixels each, covering two spectral bands -- one from 70.4 to 79.0 nm and the other from 97.3 to 104.9 nm. The slit scanning mechanism can cover up to 16 arcminutes, while the slit covers 14 arcminutes on the detector plane. The overall raster image dimensions are up to $960\times 840$ arcseconds. SPICE's observing distance varies between about 0.3 and 1.0 AU, which must be considered when comparing these numbers with Earth-based instruments.
			
			A primary design objective of SPICE is tracing the connectedness of the photosphere and low corona out to the heliosphere -- primarily by connecting abundances observed {\em in situ} by various spacecraft with those sensed remotely by SPICE. This relies on accurate identification of particular emission lines at a variety of temperatures; lines which have varying susceptibility to ionization (the FIP, or First Ionization Potential, effect) \citep[see][for more detail]{SPICEInstrument_AA2020}. An additional important feature of SPICE for diagnosing connectedness is the use of the Doppler effect to discriminate upflows from downflows and to identify potential solar wind outflow regions deep in the corona and transition region.
			
			\noindent Although SPICE is not a high resolution instrument by design, these and other observables, along with their driving science objectives, require reasonable resolution within the limits of its design \citep[given the mission's nominal requirements; see][]{SPICEInstrument_AA2020}; degraded resolution beyond the design limits will impact the ability to achieve SPICE's science goals. This paper reports on some issues that degrade SPICE's resolution, and a novel data processing methodology which will remove the degradation.

\begin{figure*}[!htbp]
				\begin{center}
					\includegraphics[width=\textwidth]{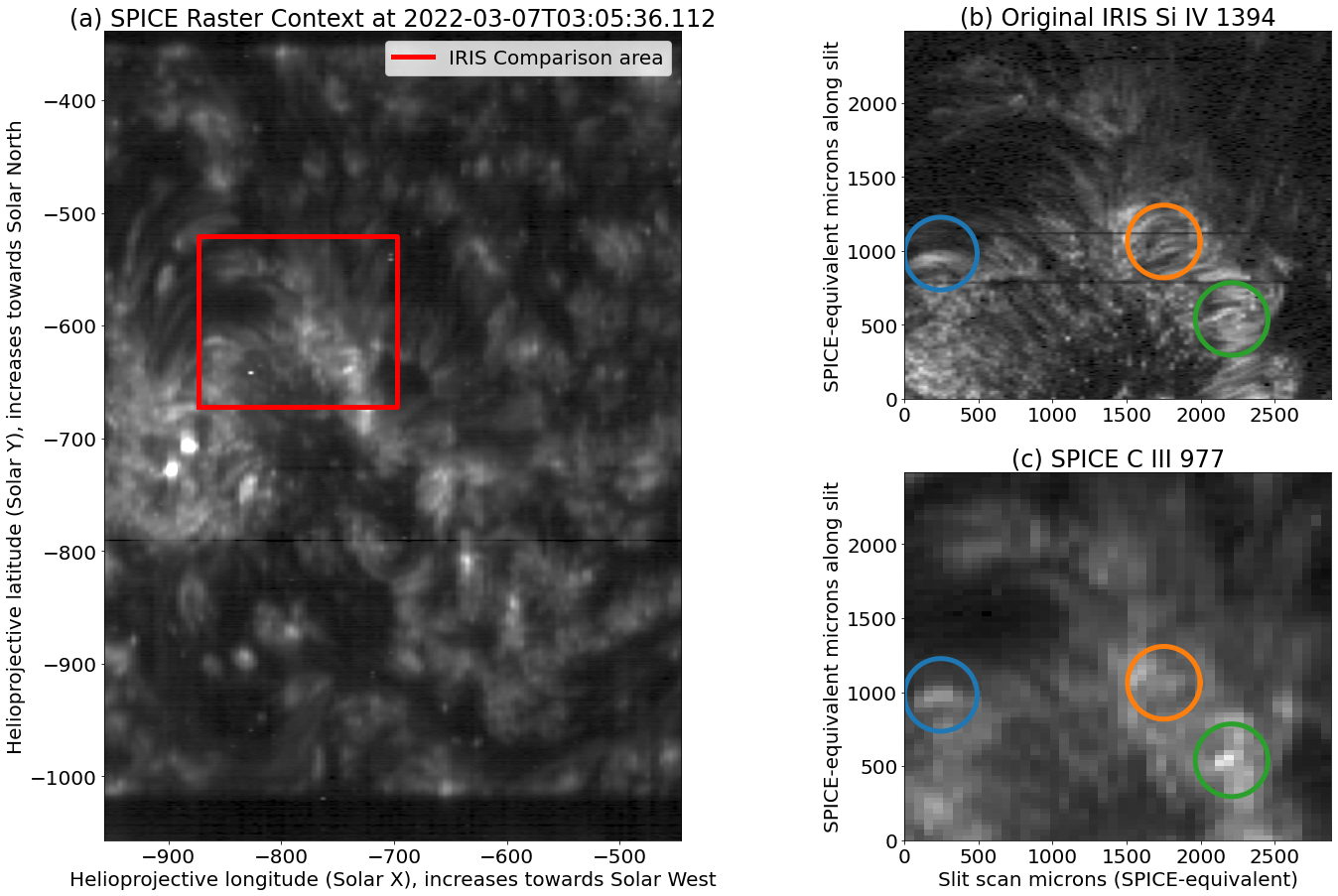}
				\end{center}
				\caption{Context image of the SPICE-IRIS coordinated observation data (a, left). This is made from one of the full-size SPICE rasters; It is a spectral sum over the C III 977 \AA\ window. The detailed comparisons in this paper will focus on the smaller region outlined in this image; spectral line fit amplitude images of this smaller region are shown in right panels, from both IRIS (upper right, b) and SPICE (lower right, c).}\label{fig:example_region_context}
			\end{figure*}

			We will be using a set of coordinated SPICE and high spectral and spatial resolution IRIS \citep[Interface Region Imaging Spectrograph][]{DePontieu_IRIS_SoPh2014} data for the examples and demonstrations in this paper, with IRIS data used as `ground truth' to assess the quality of the deconvolution and constrain the effective SPICE PSF. The observations were taken on March 7, 2022, when SPICE's orbit placed it on the Earth-Sun line, with the same perspective as IRIS. We will compare the SPICE C III 977 \AA\ observations with IRIS Si IV 1394 \AA\ observations; These  two lines form very close in terms of temperature, at log T [K] of  4.78 and 4.81 in ionisation equilibrium \citep[e.g., Table 1 of][]{PeterEtal_ApJ2006}. An image showing the larger context of this region is shown in Figure \ref {fig:example_region_context}.
			 			 
		\section{Overview of SPICE PSF Problem and Possible Solution}
		\subsection{PSF Issues and need for correction}

\begin{figure*}[!htbp]
				\begin{center}
					\includegraphics[width=\textwidth]{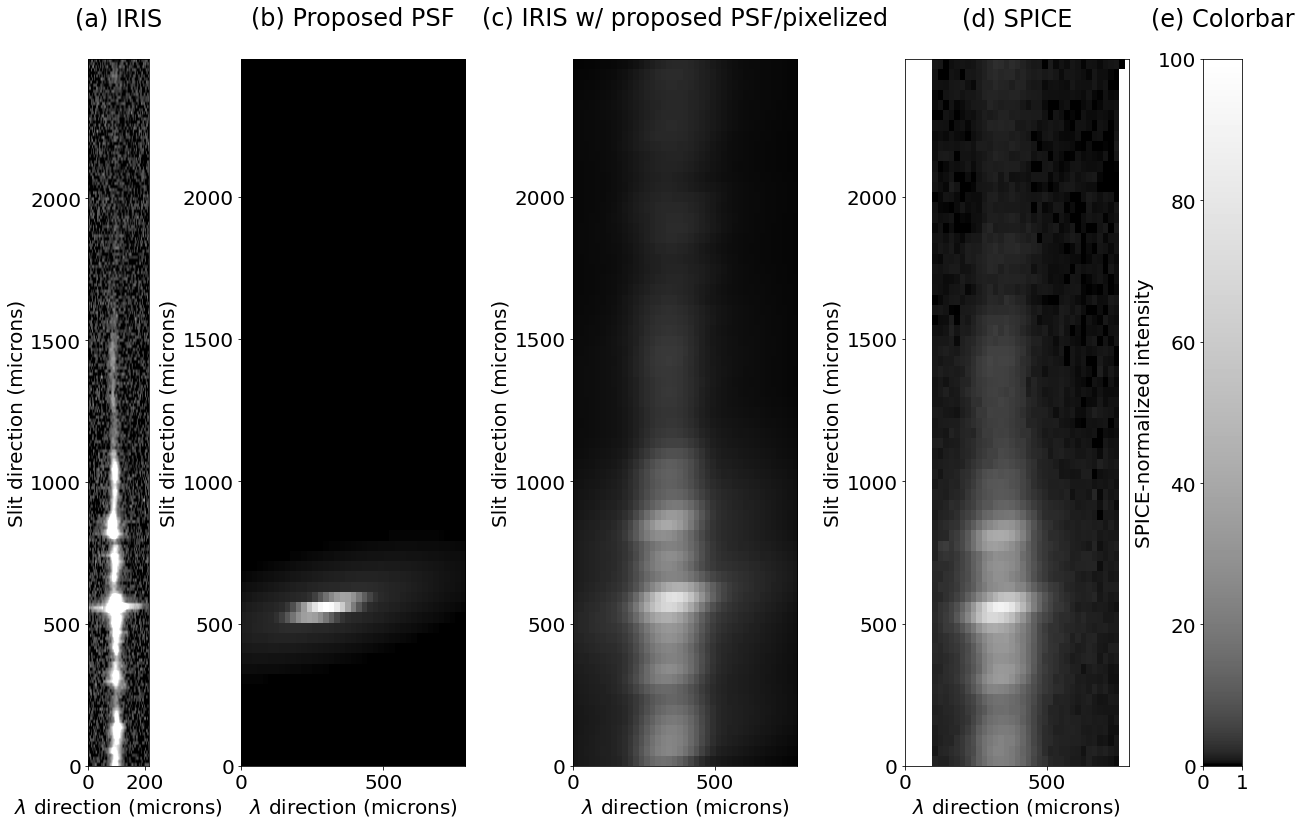}
				\end{center}
				\caption{Spectra from a particular slit position in the SPICE and IRIS data. This slit position is chosen to have good agreement between SPICE and IRIS and passes through the middle of the green circled region in Figure \ref{fig:example_region_context}. On the left (a) is the IRIS spectra, second from left (b) shows the proposed SPICE PSF with SPICE pixelization, third from left (c) is the IRIS data (`synthetic SPICE') with this PSF and pixelization, and finally (d) we show the actual observed SPICE data. The elongation and apparent rotation of the PSF are evident, as is the agreement between (synthetic) IRIS and SPICE.}\label{fig:IRIS_SPICE_spectral_scan}
			\end{figure*}

			\noindent Since SPICE is a scanning slit spectrograph, its observables are in three dimensions -- spectral ($\lambda$), slit-aligned ($y$), and slit-perpendicular ($x$, the scanning direction; actually, there is a fourth time dimension but we do not consider this here because we assume the source and PSF do not change with time). The SPICE PSF is similarly three-dimensional, extending in all three of these directions. The issue this paper addresses is caused by a PSF that is both elongated and rotated with respect to these axes, rather than being aligned with them. We have observed that the SPICE PSF is rotated both within the slit plane (i.e., $y-\lambda$) and in some cases in the scanning plane as well. This is shown for an example slit plane by Figure \ref{fig:IRIS_SPICE_spectral_scan}; this slit plane passes through the middle of the green circled region in Figure \ref{fig:example_region_context}. A rotation on the detector (i.e., $y-\lambda$) plane of $\sim 15^\circ$ is evident (this angle will appear different if the pixels are not square, and can vary with the different slit width; this paper focuses on the 2 arcsecond slit). We believe this issue is caused by a combination of astigmatism on the primary mirror and defocus.

			This elongation and rotation of the SPICE spectro-spatial PSF impacts its ability to diagnose connectedness and other important instrument capabilities in a variety of ways. It reduces the definition with which SPICE can resolve boundaries of regions with differing connectedness -- some of these can be rather small and in general small changes in the region identification can result in large impacts on the quality of the association with {\em in situ} observations. The degradation also hampers the ability to resolve the fainter spectral lines when there are neighboring bright spectral lines. This is especially critical because the selection of lines which adequately sample both abundance and temperature is very limited (the aforementioned spectral blends are a case in point).
		
			Even more challenging is that the tilt of the PSF between that spatial and spectral directions causes a bright feature at one location to create the illusion of a neighboring dim feature, which is {\em Doppler shifted} relative to the true feature -- i.e., bright features appear to have Doppler shift lobes next to them; Together with Doppler fitting, the elongated and rotated `effective PSF in the $y-\lambda$ plane acts like an edge detection filter, causing bright features to be ‘aliased’ from intensity into Doppler (As previously mentioned, we believe this issue is ultimately caused by a combination of astigmatism on the primary mirror and defocus). This confounds attempts to understand connectedness and dynamics by looking at Doppler shifts, and we have seen this kind of feature in the SPICE data.

\begin{figure*}[!htbp]
				\begin{center}
					\includegraphics[width=\textwidth]{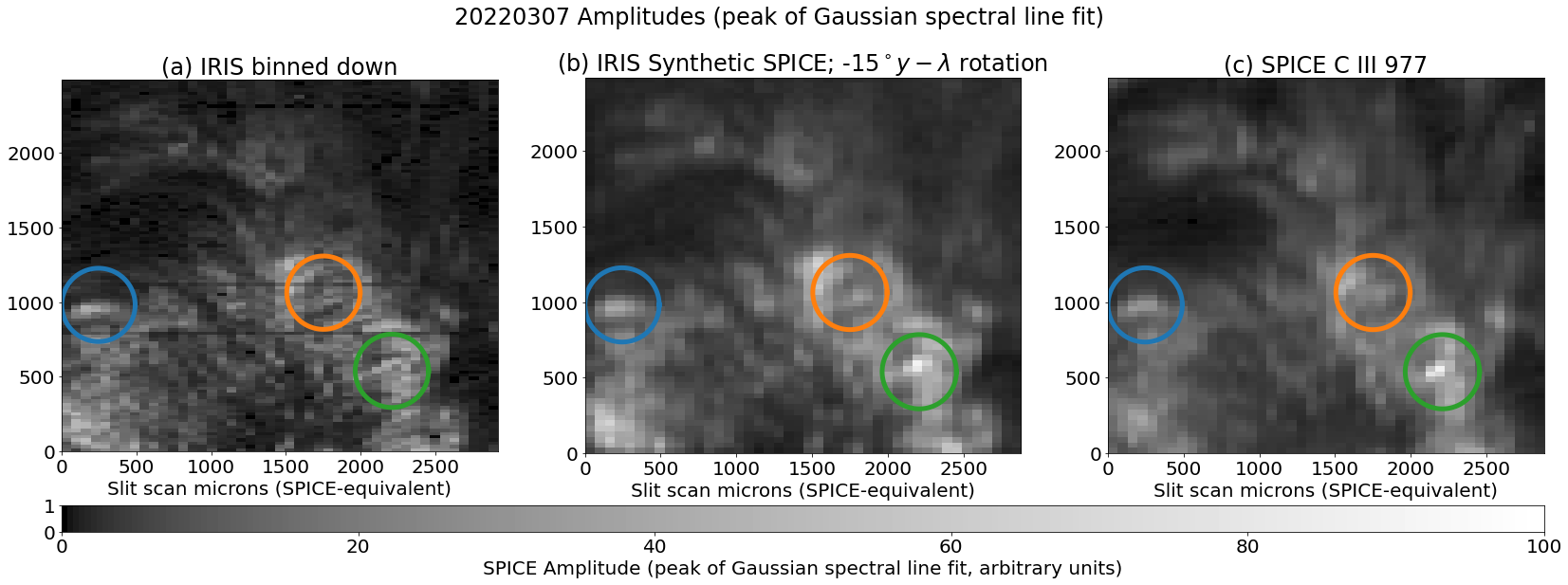}
					\includegraphics[width=\textwidth]{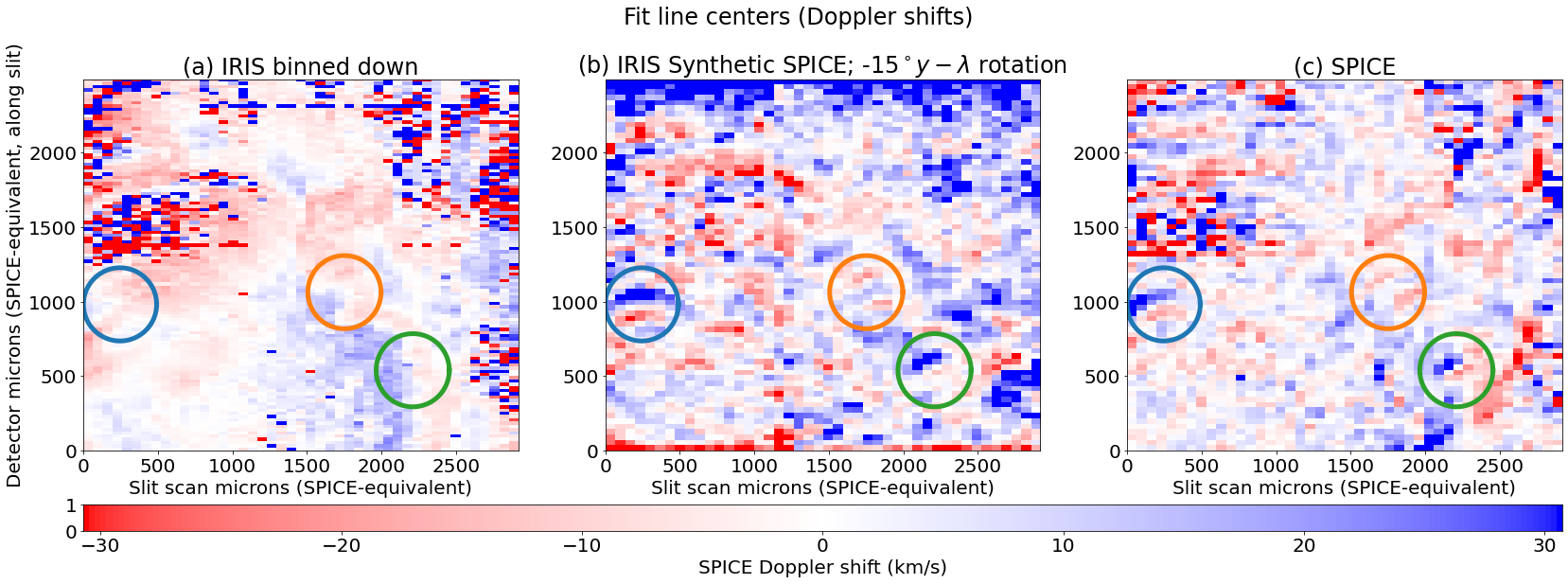}
				\end{center}
				\caption{Amplitude (upper row) and Doppler (lower row) fits to the IRIS (left column, a), IRIS with SPICE PSF and pixelization (`IRIS synthetic SPICE'; center column, b), and SPICE (right column, c). Before fitting, the IRIS data on left were first binned down $2\times 2$ to better match the SPICE pixel size. Both the `synthetic SPICE' and real SPICE Doppler (b and c) show clear ridge artifacts which are absent from the IRIS Doppler (a). Some specific artifacts are highlighted; comparison with the amplitudes in the upper row shows that these are associated with bright features in intensity. The elongated, rotated effective $y-\lambda$ PSF causes bright intensity features to alias into Doppler, as further discussed in the text.}\label{fig:example_doppler_fits}
			\end{figure*}

			To illustrate the issue, we show (in the lower panels of Figure \ref{fig:example_doppler_fits}) the Doppler fits to the binned down IRIS data, the IRIS data degraded with the SPICE PSF, and the actual SPICE data, for the region of interest already highlighted by Figure \ref{fig:example_region_context}. Strong Doppler features are evident in the IRIS data degraded with SPICE PSF and the SPICE data which are not present in the original IRIS data. We have circled three of the most prominent such features in the Doppler map. Each of these coincide with bright features (e.g., ridges) in intensity, as can be seenin the lower panels of Figure \ref{fig:example_doppler_fits}. The clear appearance of the same features in IRIS with the SPICE PSF, when they are absent from the original (plus down-binning) IRIS data, clearly implicates the PSF in creating these artifacts. A similar effect and artifacts have been previously noted for Hinode EIS \citep{HinodeEIS}. This is primarily discussed in an online EIS SW note \citep{EIS_SWNOTE8}, but also alluded to in other papers such as \cite{YoungEtal_ApJ2012}.

		\subsection{Solution method: general linear (sparse) solvers and why we're using them}
				
			These examples suffice to illustrate that this PSF effect seriously degrades the quality of Doppler data observed by SPICE. A means of removing it is highly desireable, and the objective of this work is to develop such a method which could ultimately be applied to all data at some level of its data pipeline. The conventional way of removing PSFs is via convolution \citep[e.g.,][]{PoduvalEtal_AIAPSF2013}. Convolution in this context is the translationally invariant special case of the general linear transform, and usually also assumes that the input and output of the transform have the same dimensions (e.g., $n_x$ by $n_y$ pixels). Rather than treat the problem under this special case, however, we instead treat it as an example of the full general linear transform. There are a number of reasons for this choice, including: 
		
			\begin{itemize}
				\item Pragmatically, we expect these PSF artifacts to vary over the image plane, Which a convolution cannot represent;
				\item The general transform provides a somewhat more straightforward framework for applying other constraints such as positivity and regularization, one where we have already had success in other contexts \citep{PlowmanCaspi2020, Plowman2021}.
				\item The general full linear problem is far more broadly applicable than the (de)convolution problem. Once properly developed, the same tools used for this problem with SPICE can be applied to almost every other coronal remote sensing problem, to similarly good effect. See \cite{Plowman2021} for an early example. We will return to this point at the end of the paper.
				\item The tools for tackling the inverse part of this problem have already been extensively developed by the computer science communities, in the form of high performance sparse matrix solvers. All we need to do, therefore, is implement the forward transform in the appropriate fashion, i.e., as a sparse matrix.
			\end{itemize}
		
			We turn next to our implementation of this method. The derivation shown next presupposes a 3D PSF as a function of $x$, $y$, and $\lambda$ (the most general case for SPICE). In practice, our understanding of the problem is that it can be represented by two PSFs. The first blurs in $x$ and $y$ but does not affect $\lambda$. Therefore, this PSF does not contribute to the Doppler artifacts we are trying to correct. The other `effective' $y-\lambda$ PSF is assumed to affect $y$ and $\lambda$ but not $x$. This is the PSF that is elongated in $\lambda$ and rotated $\sim 15^\circ$, and which causes the Doppler artifacts. We have done a variety of tests with 2D and 3D PSFs, and found that correcting this effective PSF suffices to remove the Doppler artifacts, working as well as the 3D correction while being much faster. We leave the derivation presented here in 3D because it is more general and was already written that way. The code for doing the correction is written to be agnostic to the dimensionality of the problem, as the supplementary 1D and 2D examples demonstrate.

	\section{Introducing our solution method (Formalism)}
		\subsection{SPICE PSF as instance of general linear forward problem}
		
			\subsubsection{The SPICE forward problem}

				To begin, let us suppose SPICE observes some static (meaning it does not change over the time of observation) cube on the sky, with spectral radiance $L(x_s, y_s, \lambda_s)$ as a function of sky longitude and latitude (which we will write as $x_s, y_s$) and wavelength $\lambda_s$. These are projected onto the detector plane by the PSF, $P(x_d, y_d, \lambda_d, x_s, y_s,\lambda_s)$, which is a function of both sky coordinates ($x_s, y_s, \lambda_s$) and detector plane coordinates ($x_d, y_d, \lambda_d$). In reality, $x_d$ and $\lambda_d$ are time-multiplexed onto the same surface (although not necessarily the same axis) by the grating, slit, and scanning mechanism, but we will treat them as independent for clarity of demonstration. The flux density $E(x_d, y_d, \lambda_d)$ on the detector plane observed for this cube is the integral of the cube against the PSF over all sky angles:
		
				\begin{equation}\label{eq:flux_density}
					E(x_d, y_d, \lambda_d) = \int P(x_d, y_d, \lambda_d, x_s, y_s,\lambda_s) L(x_s,y_s,\lambda_s)dx_s dy_s d\lambda_s
				\end{equation}
			
				Where we have made the small angle approximation and placed the Sun at the equator of the coordinate system which allows us to set the $\cos y_s$ weight in the integral to one. The sensors divide the detector plane into pixels which we will index by $i, j,$ and $k$, each of which have weight functions (typically non-overlapping `bins') $\theta_{ijk}(x_d, y_d, \lambda_d)$. The fluxes $\Phi_{ijk}$ into each of these pixels are the integrals of the detector plane flux densities against each of these weight functions:
			
				\begin{equation}
					\phi_{ijk} = \int \theta_{ijk}(x_d, y_d, \lambda_d) E(x_d, y_d, \lambda_d) dx_d dy_d d\lambda_d
				\end{equation}
			
				At this point we point out what we call the overall response function of the $\{ijk\}$th pixel of the instrument,
				\begin{equation}
					R_{ijk}(x_s,y_s,\lambda_s) = \int \theta_{ijk}(x_d,y_d,\lambda_d P(x_d, y_d, \lambda_d, x_s, y_s,\lambda_s) dx_d dy_d d\lambda_d
				\end{equation}
			
				The name of the game is to find the spectral radiance, $L$, which is an unknown continuous function and therefore has an infinite number of degrees of freedom. To limit the number of degrees of freedom we can without loss of generality (in practice if not in principal) define $L$ in terms of a linear combination of a set of appropriately chosen basis functions, $B_{lmn}(x_s, y_s, \lambda_s)$:
			
				\begin{equation}
					L(x_s, y_s, \lambda_s) = \sum_{lmn} c_{lmn} B_{lmn} (x_s, y_s, \lambda_s),
				\end{equation}
			
				where $c_{lmn}$ are the coefficients of the linear combination. The basis function set ought to be linearly independent (i.e., no basis function can be expressed as a linear combination of the others). It is useful for the present purpose if each one is spatially compact and does not overlap much (or at all) with the other basis functions. They should also cover the sky plane at least as well as the pixels cover the detector plane. A set of pixel like `bins' in $\{x_s, y_s, \lambda_s\}$ with equivalent (or denser) spacing to the pixels will suffice and is the choice we use in this paper. The indices $\{l,m,n\}$ are intended to reflect the identification of the bins with the coordinate axes. In terms of this, the pixel fluxes are

				\begin{equation}
					\phi_{ijk} = \sum_{lmn} c_{lmn} \int R_{ijk}(x_s,y_s,\lambda_s) B_{lmn}(x_s,y_s,\lambda_s)dx_s dy_s d\lambda_s.
				\end{equation}

				If we recognize an array of {\em response coupling terms},
				\begin{equation}
					r_{ijk,lmn} \equiv \int R_{ijk}(x_s,y_s,\lambda_s) B_{lmn}(x_s,y_s,\lambda_s)dx_s dy_s d\lambda_s,
				\end{equation}
				(the consituents of this are all known and computable, being composed of the basis functions and the in principal known instrument response functions) we are left with the following linear equation relating the unknowns ($c_{lmn}$) to the knowns ($\phi_{ijk}$):
				\begin{equation}
					\phi_{ijk} = \sum_{lmn} r_{ijk, lmn} c_{lmn}
				\end{equation}
				We have used three indices each for the coefficients and the observables, but this is simply a notational convenience that allows us to identify each of the coordinate axes with its own index. Mathematically, we can `flatten' all these indices down to a pair of indices (one for the coefficients, and one for the observables) by relabeling according to the following one-to-one correspondence:
				\begin{equation}\label{eq:multiplexing_eq1}
					in_jn_k + jn_k + k \rightarrow i
				\end{equation}
				\begin{equation}\label{eq:multiplexing_eq2}
					ln_mn_n + mn_n + n \rightarrow j
				\end{equation}
			
				Our linear equation then becomes
				\begin{equation}\label{eq:flattened_fluxes}
					\phi_i = \sum_j F_{ij} c_j,
				\end{equation}
				Which we recognize as a familiar matrix equation, with the array of coupling terms becoming a {\em forward response matrix}. Any N-dimensional linear discrete or integral transform can be similarly treated, and nonlinear ones may be amenable to a linearization and iteration approach, so this can be a very powerful and general method for dealing with these problems.

			\subsubsection{1D and 2D Examples}			
\begin{figure*}[!htbp]
					\begin{center}
						\includegraphics[width=\textwidth]{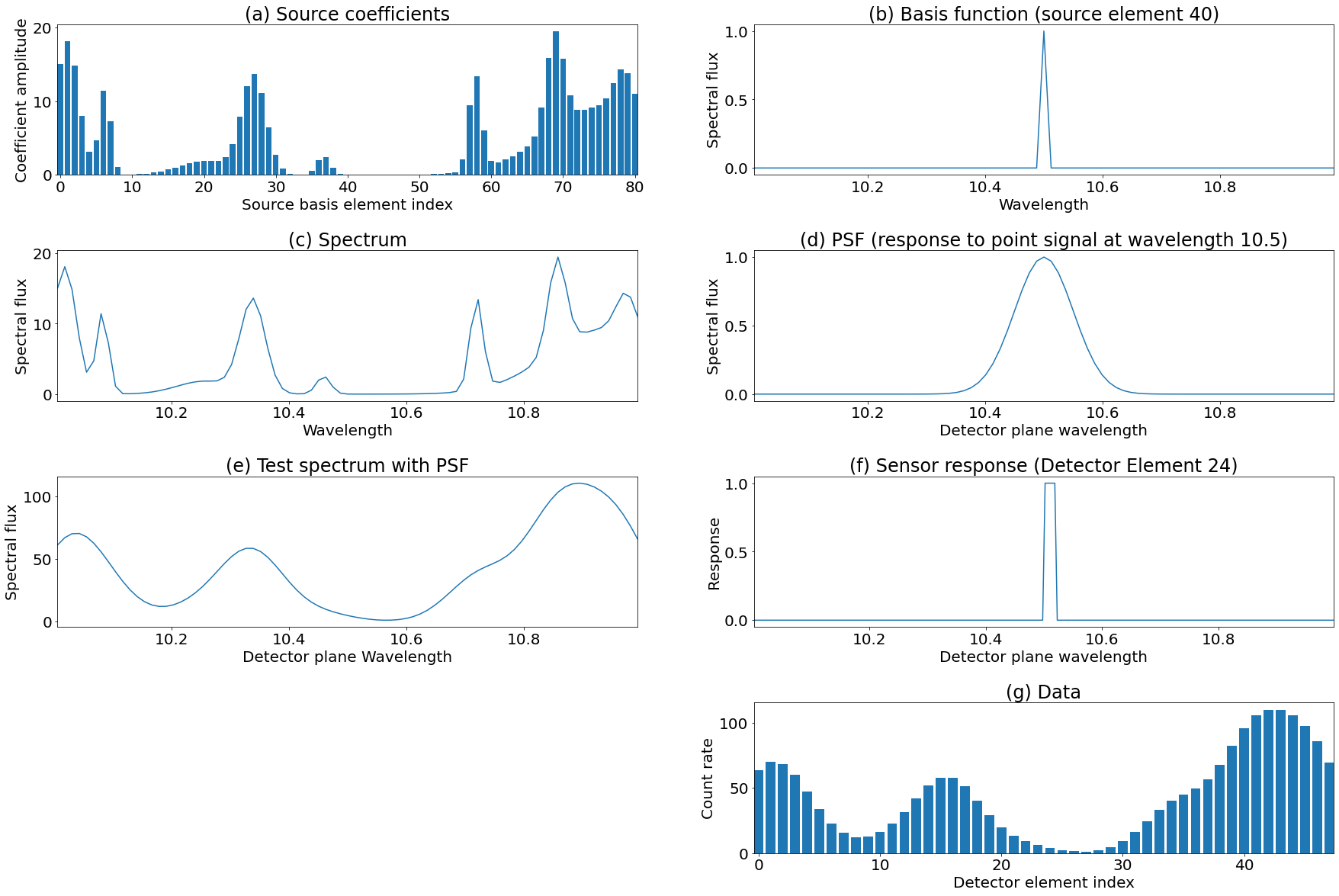}
					\end{center}
					\caption{One-dimensional example of the linear process relating a source model to observations. The input to the model (a, first row, left) is a set of coefficients. These are multiplied by a set of basis functions at various wavelengths (one is shown at right on first row, b) and the result is added to produce a spectrum (second row on left, c). The spectrum in turn is multiplied by the PSF at each wavelength (d) and the result added to produce a detector plane spectrum (3rd row on left, e). Lastly, each of the sensor response functions (3rd row on right, f) are integrated against the detector plane spectrum to produce count rates in each detector element (4th row on right, g).}\label{fig:OneD_example}
				\end{figure*}

\begin{figure*}[!htbp]
					\begin{center}
						\includegraphics[width=\textwidth]{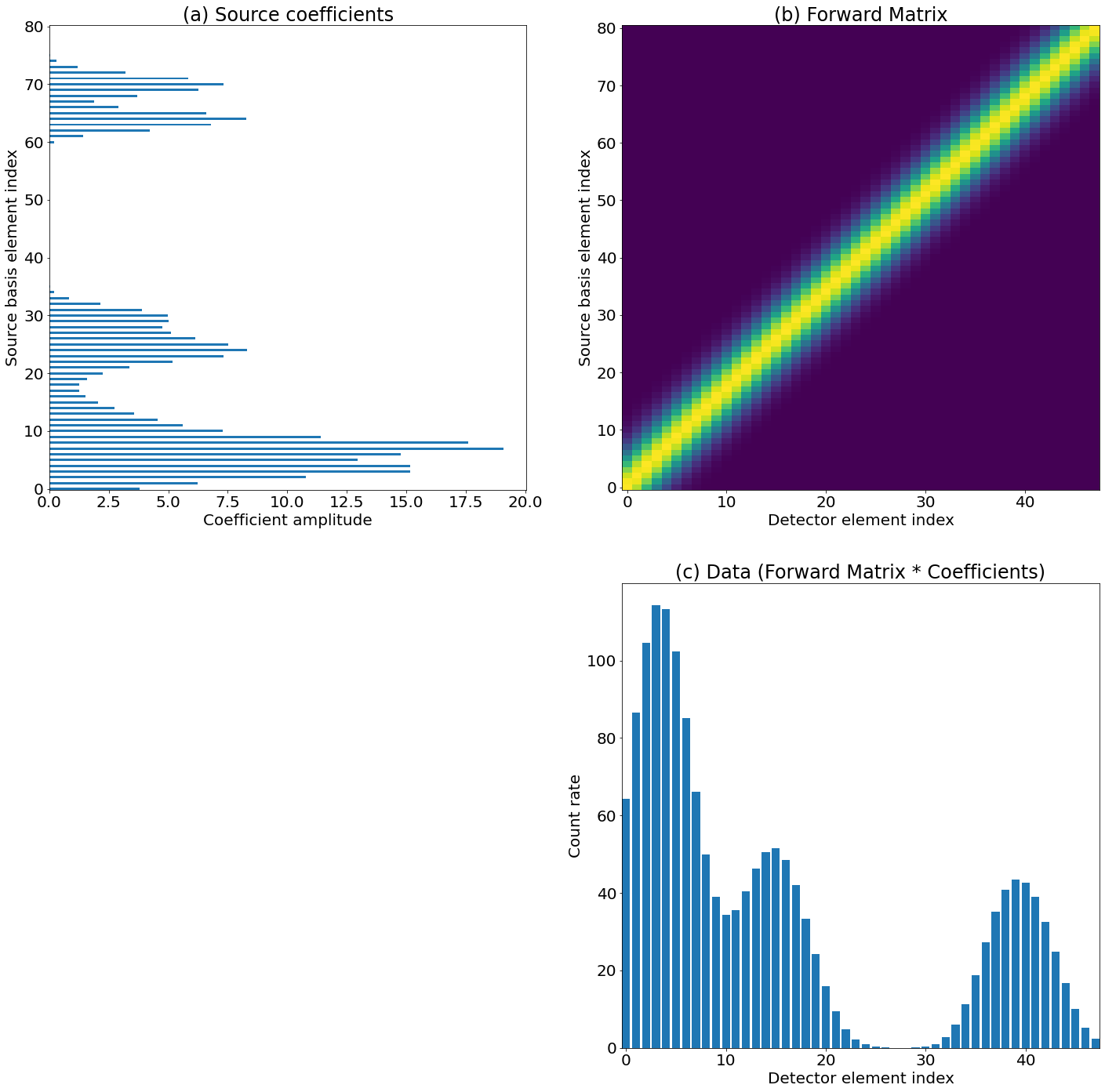}
					\end{center}
					\caption{The input (source coefficients, a at top left) and outputs (detector count rates, c at bottom right) in this figure are identical to Figure \ref{fig:TwoD_example}, but with all of the intermediate operations carried out by a single matrix (b). The vertical axis of this matrix is aligned with the source element index, while the horizontal axis is aligned with the detector element index.}\label{fig:OneD_example_matrix}
				\end{figure*}

				At this point, a pair of lower dimensional examples may be illustrative. Consider the spectrum for a single pixel, i.e., a function only of (for example) wavelength. This is equivalent to the DEM problem as presented by \cite{PlowmanCaspi2020}, the only difference being that instead of the instrument temperature response functions found there, we have a combination of the wavelength PSF and pixel bin functions:
				\begin{equation}
					R_i(\lambda_d) = \int \theta_i(\lambda_d) P(\lambda_d, \lambda_s) d\lambda_d
				\end{equation}
				The forward matrix is 
				\begin{equation}
					F_{ij} = \sum_j \int \int \theta_i(\lambda_d) P(\lambda_d, \lambda_s) d\lambda_d B_j(\lambda_s) d\lambda_s c_j
				\end{equation}
				This is illustrated graphically in Figures \ref{fig:OneD_example} and \ref{fig:OneD_example_matrix}. The only difference between them is that in Figure \ref{fig:OneD_example}, the individual steps are depicted, whereas Figure \ref{fig:OneD_example_matrix} combines all of these steps into the forward matrix -- This is possible because all of the individual steps are linear operations.

\begin{figure*}[!htbp]
					\begin{center}
						\includegraphics[width=0.7\textwidth]{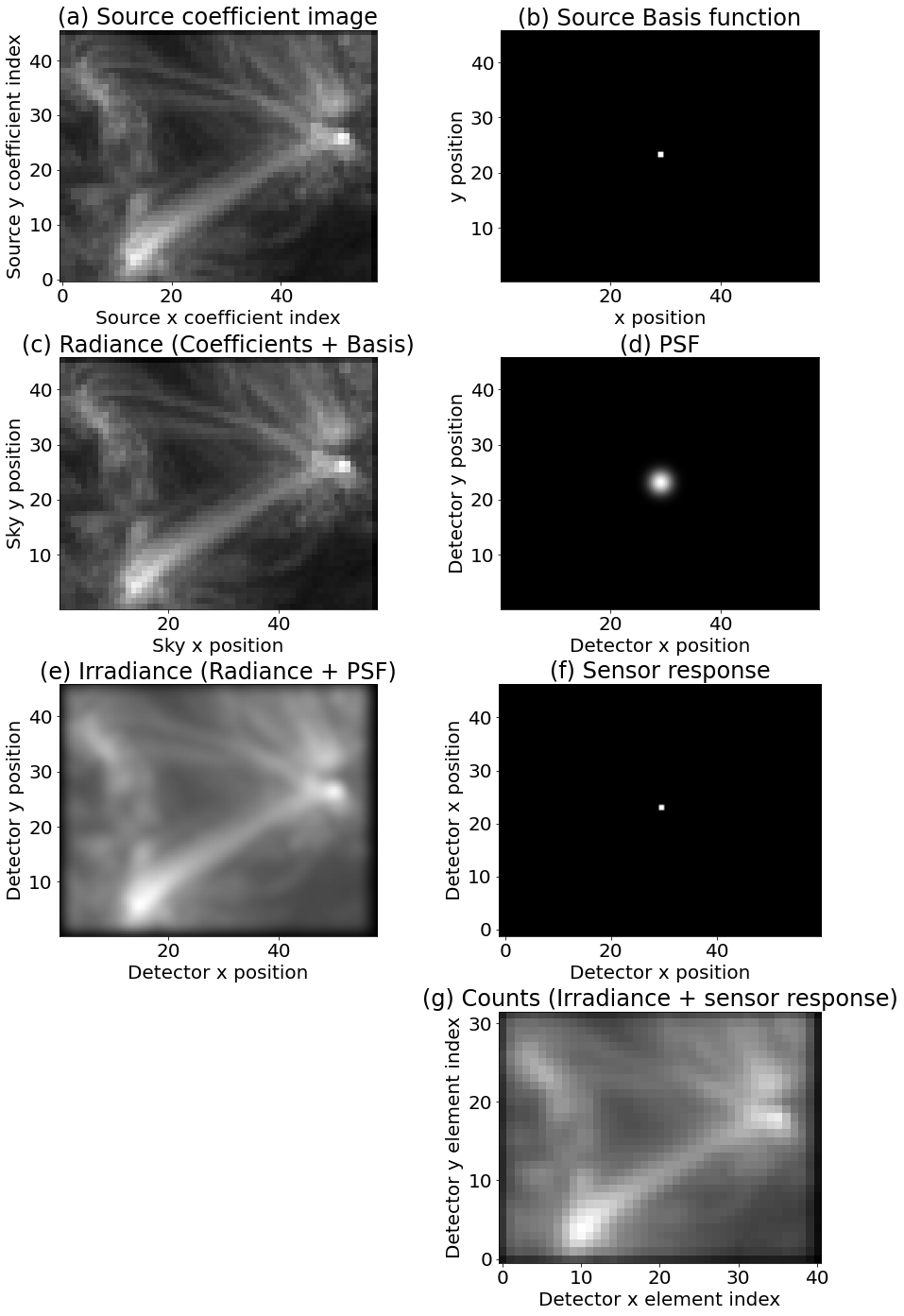}
					\end{center}
					\caption{Identical to Figure \ref{fig:OneD_example} except that in this case the source and observations are 2-Dimensional. The output resolution of these examples differs from the input, because in general the resolution of solar sources differs than that of our detectors, to demonstrate that this framework can accomodate such resolution differences.}\label{fig:TwoD_example}
				\end{figure*}

\begin{figure*}[!htbp]
					\begin{center}
						\includegraphics[width=\textwidth]{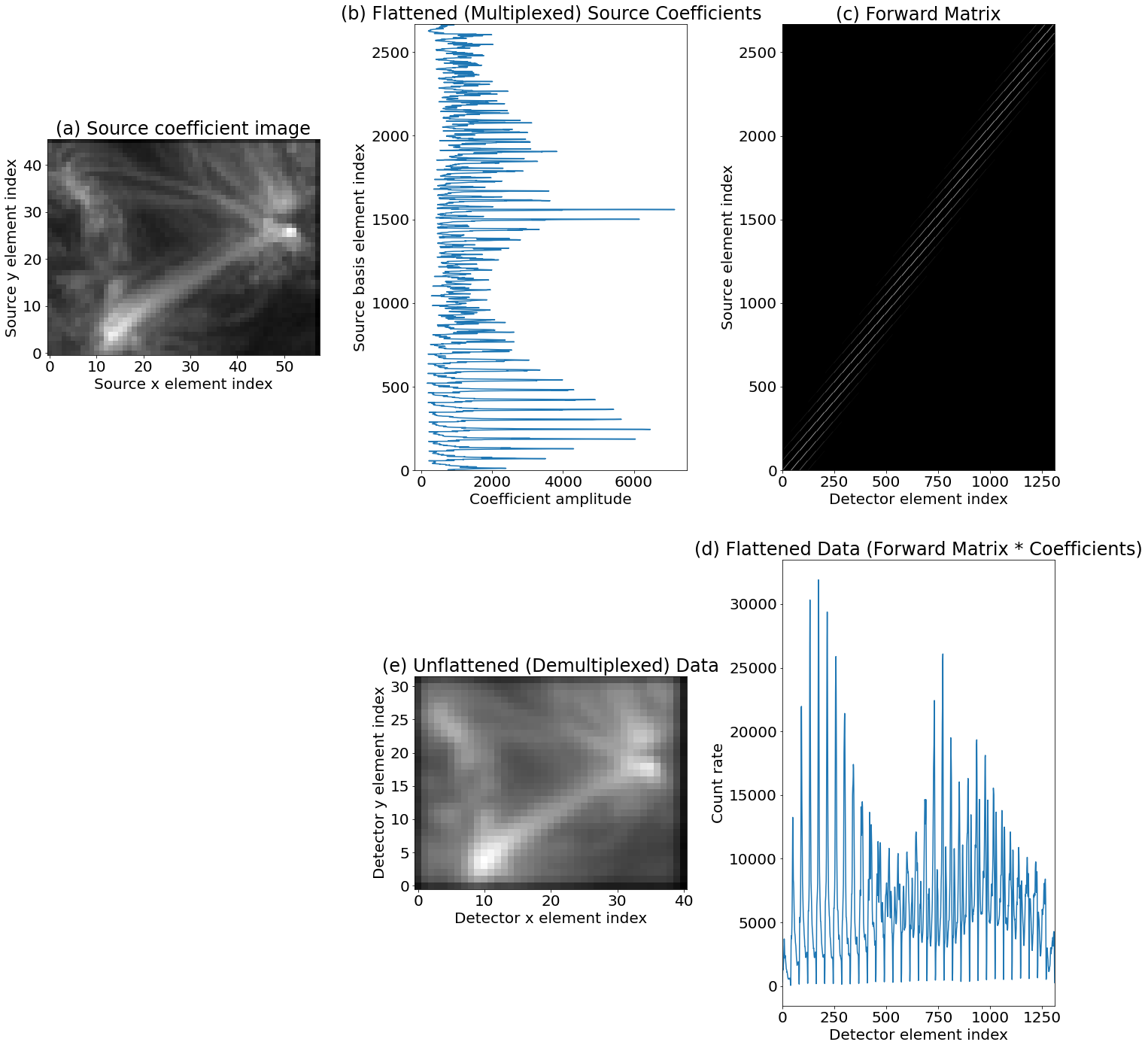}
					\end{center}
					\caption{The input (source coefficient image, a at top left) and outputs (detector count rates image, e at bottom center) in this figure are identical to Figure \ref{fig:TwoD_example}, but with all of the intermediate operations carried out by a single matrix (shown as an image at top right, c). The vertical axis of this matrix is aligned with the source element index, while the horizontal axis is aligned with the detector element index. This is equivalent to the 1D example, except that there is a multiplexing step that `flattens' the 2D input to 1D (top center, b) prior to multiplication by the forward matrix (top right). The result of this multiplication (bottom right, d) can then be demultiplexed to produce the output image (e).}\label{fig:TwoD_example_matrix}
				\end{figure*}

				Figures \ref{fig:TwoD_example} and \ref{fig:TwoD_example_matrix} are of the same format and show how the same treatment can be scaled up to two dimensions (spatial, in this example, although it makes no difference in the mathematical treatment). Essentially, the additional dimension in the source and the detector are simply multiplexed down to the one `input' (conventionally the columns) and one `output' dimension of the forward matrix. Mathematically, the only difference is the addition of a multiplexing (e.g. by `flattening' in \texttt{numpy} or \texttt{reform}ing to a 1D vector in IDL) step at the beginning and, if desired, a demultiplexing step at the end (e.g., by using \texttt{reform} in IDL or \texttt{reshape} in \texttt{numpy}) -- But these are trivial 1-to-1 operations. This additional multiplexing step is shown in the top left and bottom center panels of Figure \ref{fig:TwoD_example_matrix}, which otherwise is identical in format to the one-dimensional figure (\ref{fig:OneD_example_matrix}). The output resolution of these examples differs from the input, because in general the resolution of solar sources differs than that of our detectors, to demonstrate that this framework can accomodate such resolution differences. 

				The SPICE forward problem for the effective ($y-\lambda$) PSF is equivalent to this 2D correction, while the full forward problem follows the same trend from 1D to 2D to 3D -- the same multiplexing and demultiplexing operations suffice to convert the problem into matrix form. The practical implementation of this can be conceptually thorny, but we have developed a general N-Dimensional, coordinate aware framework in Python for computing these matrices. This will be described in more detail in Section \ref{sec:implementation}, and the framework is also available for download (see Section \ref{sec:implementation}).

			\subsubsection{Sparse Matrix Algorithms}
		
				It is clear from the examples provided above that the forward matrix is very large, even for images of modest size. However, only detector elements within PSF range of the source elements have non-zero entries, and the PSF is much smaller than the overall image. Therefore, this matrix is sparse. A wide variety of specialized sparse matrix algorithms exist to work with such matrices and they drastically reduce the real-world memory and processing requirements of working with these matrices. Standard examples can be found in, for instance, \cite{nr}. Additionally, solving linear problems based on sparse matrices is well developed field in computer science, and by casting the problem in this form we can avail ourselves of the tools that have been developed. Our framework for computing the forward matrices is built on these algorithms, specifically making use of the \texttt{scipy.sparse} sparse matrix package.
		
		\subsection{Solving the SPICE forward problem}

			The fundamentals of this solution method have also been recently discussed in \cite{PlowmanCaspi2020, Plowman2021}, but we will also cover them here to show how they apply to this case. We begin by considering the forward problem in the linear least squares framework \citep{nr}:
			
			We seek to find the `best fit' coefficient vector ($c_j$ in Equation \ref{eq:flattened_fluxes}) that minimizes the sum of squares (hence, the `least squares') of the deviation of our model prediction for the data -- $\Phi_i$ from Equation \ref{eq:flattened_fluxes} -- with the data, $d_i$. These squared deviations are weighted by the inverse of the square of the errors in the data, $\sigma_i$. This weighted sum is also known as a $\chi^2$ statistic:
			
			\begin{equation}\label{eq:modelchi2}
				\chi^2 = \sum_i \frac{(\phi_i-d_i)^2}{\sigma_i^2}
			\end{equation}  
			The above assumes that the errors in the data are normally distributed, which can be seen by noting that in that case the log likelihood is $-\chi^2$. Because this merit function is convex, it has only one minimum which occurs where the gradient of $\chi^2$ with respect to the coefficient vector is zero. Better yet, because $\chi^2$ is quadratic, its derivative is linear. The best fit coefficient vector is therefore found by solving the following matrix equation \citep[cf. the chapter in][on Linear Least Squares]{nr}:
			
			\begin{equation}\label{eq:chi2solve}
				A^T\cdot {\bf b} = A^T \cdot A \cdot {\bf c}, \quad \mathrm{or} 
			\end{equation}
			Where 
			\begin{equation}	
				A_{jk} = \frac{F_{jk}}{\sigma_j}, \quad \mathrm{ and} \quad b_j = \frac{d_j}{\sigma_j}.
			\end{equation}
			This is the standard form for matrix problems in linear algebra, and in principle the linear least squares best fit solution is
			\begin{equation}
				{\bf c} = {[A^T A]}^{-1}\cdot{A^T \cdot {\bf b}}
			\end{equation}
			Reality is never quite so simple, of course -- we need to apply some additional constraints to ensure a well-posed solution, which we discuss next.
			
			\subsection{Positivity Constraint and Regularization}
				The $A^T A$ matrix is likely to be poorly conditioned even if it is not outright singular. The blurring caused by the PSF means that details we would like to resolve are degraded by the forward transform -- it loses information -- even if the number of data points is equal to the number of coefficients; in this case, $A^T A$ will be poorly conditioned, and detector noise will lead to instabilities in the reconstruction. If the number of data points is less than the number of coefficients, then $A^T A$ will be singular and simply inverting Equation \ref{eq:chi2solve} will not be possible. To resolve this issue, we apply a positivity constraint and regularization.

				\subsubsection{Regularization}
				
					Regularization typically takes the form of an additional term to minimize which is added to $\chi^2$. In this derivation, we will take this to be of the form
					\begin{equation}\label{eq:regularization}
						\lambda \sum_{ij} c_i X_{ij} c_j.
					\end{equation}
					In other words, instead of minimizing $\chi^2$ by itself, we seek to minimize the merit function
					\begin{equation}
						M \equiv \chi^2 + \lambda \sum_{ij} c_i X_{ij} c_j,
					\end{equation}

					Where $\mathbf{X}$ is a matrix specifying the regularization. This can be as simple as the identity matrix, in which case the regularization makes the solver prefer solutions with a smaller $l^2$ norm (i.e., ${\bf c}\cdot {\bf c}$). Exactly how much smaller depends on the size of the $\lambda$ parameter, which tunes the strength of the regularization. In \cite{PlowmanCaspi2020}, the operator was composed of the derivatives (WRT temperature, in that case) of the basis functions integrated against each of the other basis functions. This results in a regularization operator that makes the solver prefer solutions with the smaller derivatives -- specifically the log of the derivative in that case. 
					
					The addition of the regularization term changes the matrix equation we wish to solve from Equation \ref{eq:chi2solve} to
					\begin{equation}\label{eq:regsolve}
						A^T\cdot {\bf b} = A^T \cdot A \cdot {\bf c} + \lambda X \cdot {\bf c}.
					\end{equation}
					
					In other words, the matrix that needs to be inverted simply has the regularization matrix times the tunable paramater ($\lambda X$) added to it. 
					
					For this class of problem, $\lambda$ is often chosen to make each the residuals of the fit to be some number $\chi_0$ (typically equal to the number of data points, or `reduced' $\chi^2$ of order unity). Noting that the residuals are ${\bf b} - A\cdot {\bf c}$, and with a slight rearrangment of Equation \ref{eq:regsolve}, this requirement is equivalent to
					\begin{equation}
						\lambda \sum_j X_{ij} c_j - \chi_0 \sum_j \frac{F_{ji}}{\sigma_j} = 0.
					\end{equation}
					We want to solve this for $\lambda$. But this is just one parameter and can't satisfy each component of this vector equation. Instead, minimize the RMS of the LHS of this equation over lambda. The result is
					\begin{equation}
						\lambda = \chi_0 \frac{\sum_i (\sum_j X_{ij} c_j) (\sum_j F_{ij}/\sigma_j)}{\sum_i (\sum_j X_{ij} c_j)^2}.
					\end{equation}

					But we don't know ${\bf c}$, either. We therefore use an initial guess for ${\bf c}$ to determining our $\lambda$. We'll need one later on when we introduce the positivity constraint, anyway. 
					
					It turns out that a simple initial guess can be formed using $F^T\cdot{\bf d}$. An analogy can be made to upscaling an image by an integer factor. More generally, the results of this procedure are roughly a model which has been subjected to the source response function twice. It will be much smoother, but features will be in the right place. This is desirable for regularization and it is more conservative and therefore avoids overfitting. It does result in an over-smoothed solution as well, however, the amount of which depends on how much sharper the source is than the results of the instrument resolution degradation. To mitigate this, we apply a final correction factor to $\lambda$. The SPICE sources are significantly sharper than the SPICE PSF, so the $\lambda$ used is about 0.1.
					
					The amplitude needs to be rescaled (since $F$ is not normalized). We choose this scaling factor to be the one that minimizes the $\chi^2$ of this initial guess compared with the data, resulting in the following initial guess for ${\bf c}$:
					\begin{equation}\label{eq:initialguess}
						{\bf c}^0 \equiv F_T\cdot{\bf d} \Big[\frac{{\bf d} \cdot A \cdot A^T \cdot {\bf d}}{\vert A \cdot F^T \cdot {\bf d}\vert^2}\Big]
					\end{equation}

					The framework as currently constructed uses regularization based the $l^2$ norm. The derivative-based regularization used in \cite{PlowmanCaspi2020} could also be done here, but it would require constructing gradient operator matrices with respect to image position and wavelength, which is doable but too involved to include in the present publication. It will be implemented in future work, however, which will open up a variety of possibilities for including more physical constraints on inversions using this framework, as well as some other intrinsic benefits.
									
				\subsubsection{Positivity Constraint: nonlinear mapping}
					One condition that the sources of these data must satisfy is positivity: spectral radiances cannot be negative. This is a nonlinear constraint, which can't be directly captured by a linear formalism. Instead, we use the same nonlinear mapping on the coefficients as in \cite{PlowmanCaspi2020}. We express $c_j$ in terms of a second set of parameters, $s_j$, such that
					\begin{equation}
						c_j = g(s_j).
					\end{equation}
					We then choose the mapping function, $g$, to be one whose range is limited to the non-negative numbers. Two such examples are $e^{s_j}$ and $s_j^2$; we use the former (exponential) mapping here, as it seems to produce better convergence properties. We then linearize and iterate beginning from an initial guess. The results of this are described in more detail in \cite{PlowmanCaspi2020}, and we will not repeat all of the steps of the derivation here although we have changed the notation somewhat in an attempt at better clarity. The one difference is that in \cite{PlowmanCaspi2020} only the case where the regularization acts on the $s_j$, rather than the $c_j$ was considered. If the regularization acts on the $c_j$ instead, there are additional terms (shown below). 
					
					The result is the following matrix equation
					
					\begin{equation}
						\sum_j \alpha_{ij} c_j = \beta_i, \quad \mathrm{where}
					\end{equation}
					
					\begin{equation}
						\alpha_{ij} \equiv g'(s_i^0) \sum_k \frac{F_{ik} F_{jk}}{\sigma_i^2}g'(s_j^0) + h'(s^i_0) X_{ij} h'(s^j_0),
					\end{equation}
					and
					\begin{eqnarray}\nonumber
						\beta_i &\equiv& \sum_j g'(s_i^0)\frac{F_{ji}}{\sigma_j^2}\big[d_j - \sum_k F_{jk} (g(s_k^0)-s_k^0g'(s_k^0))\big] \\
						&& - h'(s_i^0)X_{ij}\big[h(s_j^0)-s_j^0h'(s_j^0)\big]
					\end{eqnarray}
					
					\noindent where $h(s_i)$ is the mapping function for the regularization. It is equal to $g(s_i)$ if the regularization is with respect to ${\bf c}$ and $h(s_i)=s_i$ if it is with respect to ${\bf s}$, and primes on $g$ or $h$ represent their derivatives with respect to their arguments. $s_j^0$ represents the coefficient parameters from the previous step of the iteration. There's no reason the regularization couldn't use some other mapping function of $s$, but we do not further consider that possibility here; we use the same mapping for both $\chi^2$ and the regularization ($h(x)=g(x)$). The regularization strength, $\lambda$, likewise changes to
					
					\begin{equation}
						\lambda = \chi_0 \frac{\sum_i (h'(s_i^0) \sum_j X_{ij} h(s_j^0)) (\sum_j g'(s_i^0) F_{ji}/\sigma_j)}{\sum_i (h'(s_i^0) \sum_j X_{ij} h(s_j^0))^2}.
					\end{equation}
					
					These equations suffice to ensure positivity, which in addition to being required by the physics makes the solutions more well-posed in many situations. It does require an iteration, but the convergence is generally good provided the mapping functions are monotonic (this makes the overall optimization problem convex, so the figure of merit has a single global minimum).
				
				\subsubsection{Solvers: GMRes, Bicstab, etc}
					Since these matrices are very large, they require sparse solvers. These generally rely on some sort of iterative process in order to avoid dealing with the fully populated matrix. The code for applying these corrections is developed in Python, so we have been using the solver algorithms included in the \texttt{sparse} package of SciPy \citep{2020SciPy-NMeth}. Specifically the stabilized biconjugate gradient \citep[\texttt{bicgstab}][]{bicgstab} and \texttt{lgmres} \citep{lgmres} algorithms. Both of these algorithms have a long heritage -- they are variants of algorithms mentioned in the 1992 edition of \cite{nr}, namely the Biconjugate Gradient and Generalized Minimal Residual methods. We have tested both and found their convergence and performance to be comparable and more than acceptable for the problem at hand. 

		\subsection{Implementation Details \& Practical Considerations}\label{sec:implementation}
	
			To accompany this formalism, we have developed a flexible coordinate-aware Python framework which can compute basis functions, PSF, and response matrix for arbitrary source and detector dimensions. It contains the following components:
			\begin{description}
				\item[Basis functions] Rectilinear N-Dimensional `bin' and `triangle' functions are provided as examples, and others can easily be added.
				\item[Point Spread Functions] Functions specifying how an N-Dimensional detector plane is illuminated by a point source at a particular location. A Gaussian-based `boilerplate' PSF is included as an example (see Section \ref{sec:boilerplatepsf} below).
				\item[Coordinate Grids] These represent the standard notion of an N-Dimensional array of coordinate points indexed to a grid. They are defined in terms of a coordinate system and a transform from grid indices to a set of coordinate points. The transform may be defined in terms of an origin (the location of the center of the element with all indices zero), a set of sizes ($\Delta_x$, $\Delta_y$, etc), and the dimensions of the array, but more complex transforms are also possible. This transform must be one-to-one.
				\item[Source and Detector Element Grids] These represent the standard notion of a gridded array of elements, and can be either a set of source basis elements or detector elements (pixels). They are defined in terms of the coordinate grids and registered to it, but add code for returning the coordinates and amplitude of a particular basis element and for returning the response of all detector elements to a particular point source. Sources and detectors are independent and each have their own coordinate system.
				\item[Coordinate Transform] A transformation from the coordinate system of the source model to the coordinate system of the detector. This is build to interface with the coordinate grids, and is also setup with the \texttt{astropy} \citep{AstroPyI_AA2013, AstroPyII_AJ2018} and \texttt{sunpy} \citep{sunpy_community2020} coordinate systems in mind, but for these examples we use a trivial identity transform, leaving the source defined in terms of detector units, and registered to the same grid. This transform does not need to be one-to-one -- e.g., it can map from a set of 3D spatial points on the sky to a 2D detector plane.
			\end{description}
			This framework should be able to accommodate detectors whose elements are not gridded in the usual way -- RHESSI \citep{RHESSI} could be an example of this -- but we do not further explore the possibility in this paper. Likewise, we will leave detailed description of the framework to its codebase and example usage software, which will be made available along with the published paper.
			
		\subsubsection{`Boilerplate' PSF Model}\label{sec:boilerplatepsf}
		For the work shown next, we use a `Boilerplate' model of the PSF which consists of a two component, 2D Gaussian where each component can be raised by an additional exponent to weight it more toward its wings or its core. We found that setting this exponent to $\sim 1.5$ made for a closer match between the IRIS line profiles (with our PSF added) and the SPICE line profiles, although the effect on the corrected Doppler shifts was very small. The mathematical expression for these Gaussians is:
		\begin{equation}
			P(\Delta\mathbf{r}) = \exp\Big[-\frac{1}{2}\big(\Delta \mathbf{r}^T \Sigma^{-1} \Delta \mathbf{r}\big)^\gamma\Big],
		\end{equation}
		where $\Delta \mathbf{r}$ is the difference between the coordinate vector of the source and the coordinate vector on the image plane, e.g., $\{x_d-x_s, y_d-y_s, \lambda_d-\lambda_s\}$ in equation \ref{eq:flux_density}. $\Sigma$ is the covariance matrix of the Gaussian, and $\gamma$ is the exponent (an exponent of one is a standard Gaussian). We specify this in terms of a set of initial axis lengths (the uncertainties of the principal axes without rotation) and rotation angles, in which case one begins with a diagonal covariance matrix that has the principal axis uncertainties on the diagonals and then rotates the matrix to its final position with a succession of rotation matrice(s) -- one in the 2D case, 3 in the 3D case -- Our implementation includes an example implementation of this PSF that works in both 2D and 3D.
			
	\section{Testing \& Application}

\begin{figure*}[!htbp]
				\begin{center}
					\includegraphics[width=\textwidth]{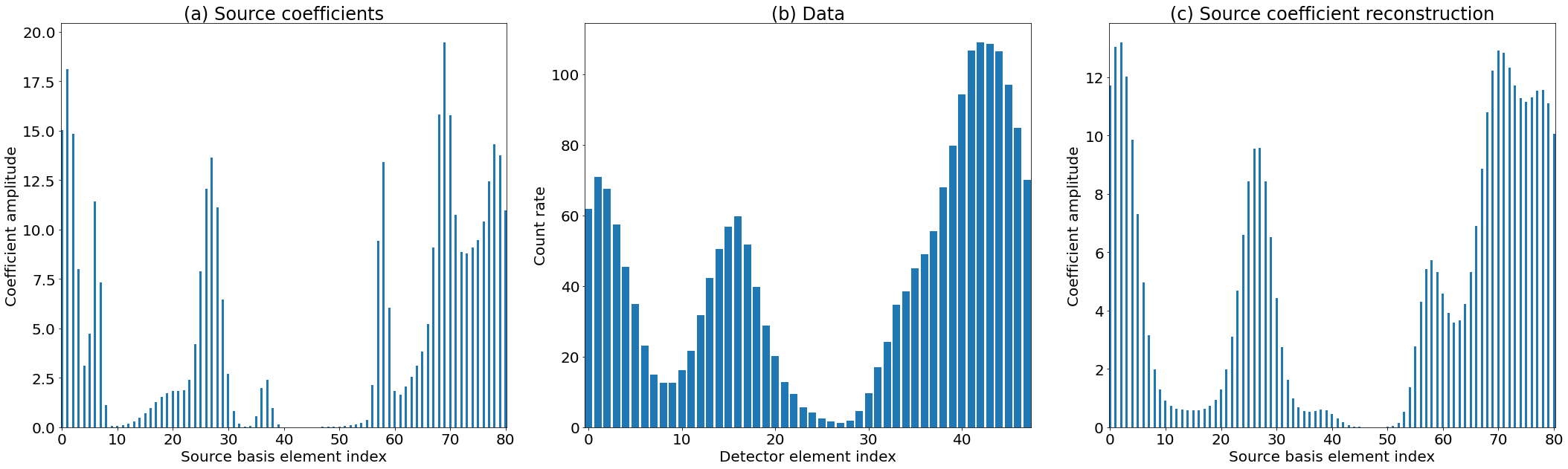}
				\end{center}
				\caption{Reconstruction of the one-dimensional example case previously shown in Figure \ref{fig:OneD_example} using our method. Left (a) is the original `source' model, center (b) is the data corresponding to that model (including PSF, pixelization, and noise), right (c) is the reconstruction of the source model.}\label{fig:example_oned_reconstruction}
			\end{figure*}

\begin{figure*}[!htbp]
				\begin{center}
					\includegraphics[width=\textwidth]{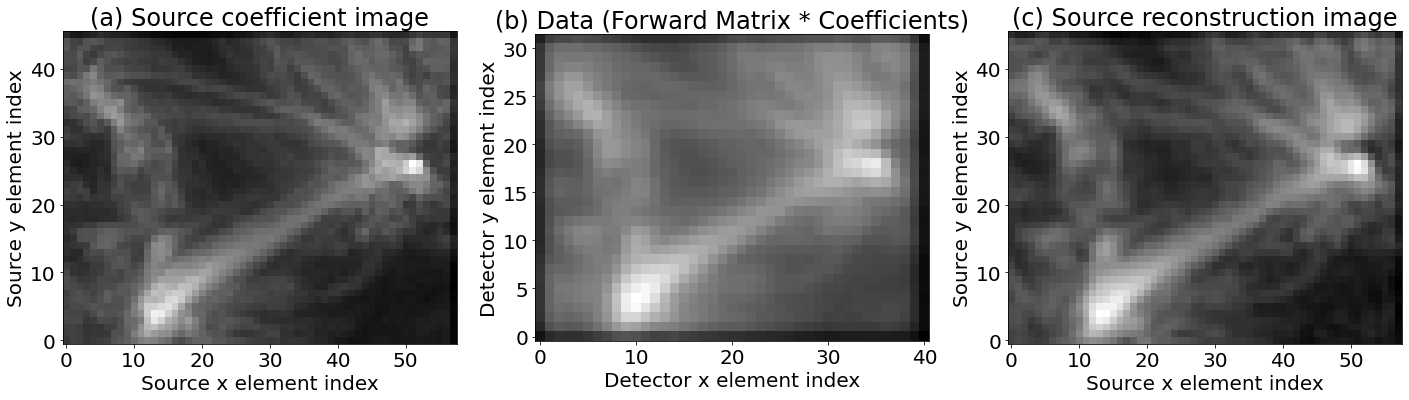}
				\end{center}
				\caption{Reconstruction of the two-dimensional example case previously shown in Figure \ref{fig:TwoD_example} using our method. Left (a) is the original `source' model, center (b) is the data corresponding to that model (including PSF, pixelization, and noise), right (c) is the reconstruction of the source model.}\label{fig:example_twod_reconstruction}
			\end{figure*}

			To begin, we show the method applied to the one and two-dimensional examples shown earlier. Figure \ref{fig:example_oned_reconstruction} shows reconstruction of the example shown in Figure \ref{fig:OneD_example}, while Figure \ref{fig:example_twod_reconstruction} shows reconstruction of the example in Figure \ref{fig:TwoD_example}. In both cases, the reconstruction recovers most of the features of the source which are not evident in the data, despite the presence of PSF, noise, and (in the 2D example) a lower pixel resolution. Apparent noise is amplified by this process, but this is an inevitable consequence of the flow of information in this problem: below some threshold, noise and features cannot be distinguished.

		\subsection{Application to SPICE data and comparison with coordinated IRIS data}

		We now show application to the example SPICE data shown earlier. As previously mentioned, we are treating the overall SPICE spatio-spectral response function as an $x-y$ response function representing (primarily) the main (spatial focusing) mirror followed by a $y-\lambda$ `effective' response function representing the elements that contribute to the spectral PSF, including the grating, and final pixelization. To reduce the size of the inverse problem, this example incorporates the downsampling in $y$, including a 2-pixel binning in that direction, into the $x-y$ response function. This is not strictly required, although it does also help to increase the SNR. The $\lambda$ resolution change, of course, all happens as part of the response function. 
		
		The correction uses the full IRIS resolution in $\lambda$; even though this is considerably higher than the SPICE resolution, it is necessary to properly resolve the spectral line -- The combination of positivity constraint, regularization, and the fact that the MTF of the PSF doesn't completely go to zero means that the correction can somewhat exceed the standard Airy/Nyquist criteria, especially since the spectral direction is mostly composed of near zero-amplitude signal. This super resolution can result in some ringing and overfitting features due to noise, however, so we apply a new forward matrix that applies a `nominal'-sized SPICE PSF and returns the pixel scale to that of the original data. The size of this `nominal' PSF is based on the resolutions described in the SPICE instrument paper \citep{SPICEInstrument_AA2020}. The parameters of the PSF used are shown in Table \ref{tab:psfparameters}, along with the slightly different parameters (just the PSF angle in this case) used for the correction of the region discussed in Section \ref{sec:correction_2ndregion}.
		
		\begin{table*}[!htbp]
			\begin{center}
				\begin{tabular}{ | l | l | l | p{7cm} |}
				\hline
				Parameter & Region 1 Value & Region 2 Value & Description \\
				\hline
				$F_{y0, c}$ & 2.0 Arcseconds & 2.0 Arcseconds & PSF core y axis FWHM before rotation or any non-Gaussian exponentiation. \\
				\hline
				$F_{\lambda 0, c}$ & 0.95 \AA\ & 0.95 \AA\ & PSF core $\lambda$ axis FWHM before rotation or any non-Gaussian exponentiation. \\
				\hline
				$\theta_c$ & $15^\circ$ & $20^\circ$ & PSF core rotation angle in degrees. \\
				\hline
				$\gamma_c$ & 1.5 & 1.5 & PSF core non-Gaussian exponent. \\
				\hline
				$F_{y0, w}$ & 10.0 Arcsrconds & 10.0 Arcseconds & PSF wing y axis FWHM before rotation or any non-Gaussian exponentiation. \\
				\hline
				$F_{\lambda 0, w}$ & 2.5 \AA\ & 2.5 \AA\ & PSF wing $\lambda$ axis FWHM before rotation or any non-Gaussian exponentiation. \\
				\hline
				$\gamma_w$ & 1.0 & 1.0 & PSF wing non-Gaussian exponent (1 means Gaussian). \\
				\hline
				$w_w$ & 0.2 & 0.2 & Wing weight (core weight is $1.0-w_w$). \\
				\hline
				\end{tabular}
				\caption{Parameters of the PSF used for the correction of the two regions; Region 2 is the second region, discussed in Section \ref{sec:correction_2ndregion} (Region 1 is shown in Figures \ref{fig:example_region_context} and Region 2 in \ref{fig:context_image_region2}). PSF size parameters must be converted to natural detector plane units (microns) before the PSF is evaluated, else they're dimensionally inconsistent. The conversion factors are $18/1.1$ microns per arcsecond and 200 microns per \AA\ (i.e., \AA\ in wavelength).}\label{tab:psfparameters}
			\end{center}
		\end{table*}

\begin{figure*}[!htbp]
			\begin{center}
				\includegraphics[width=\textwidth]{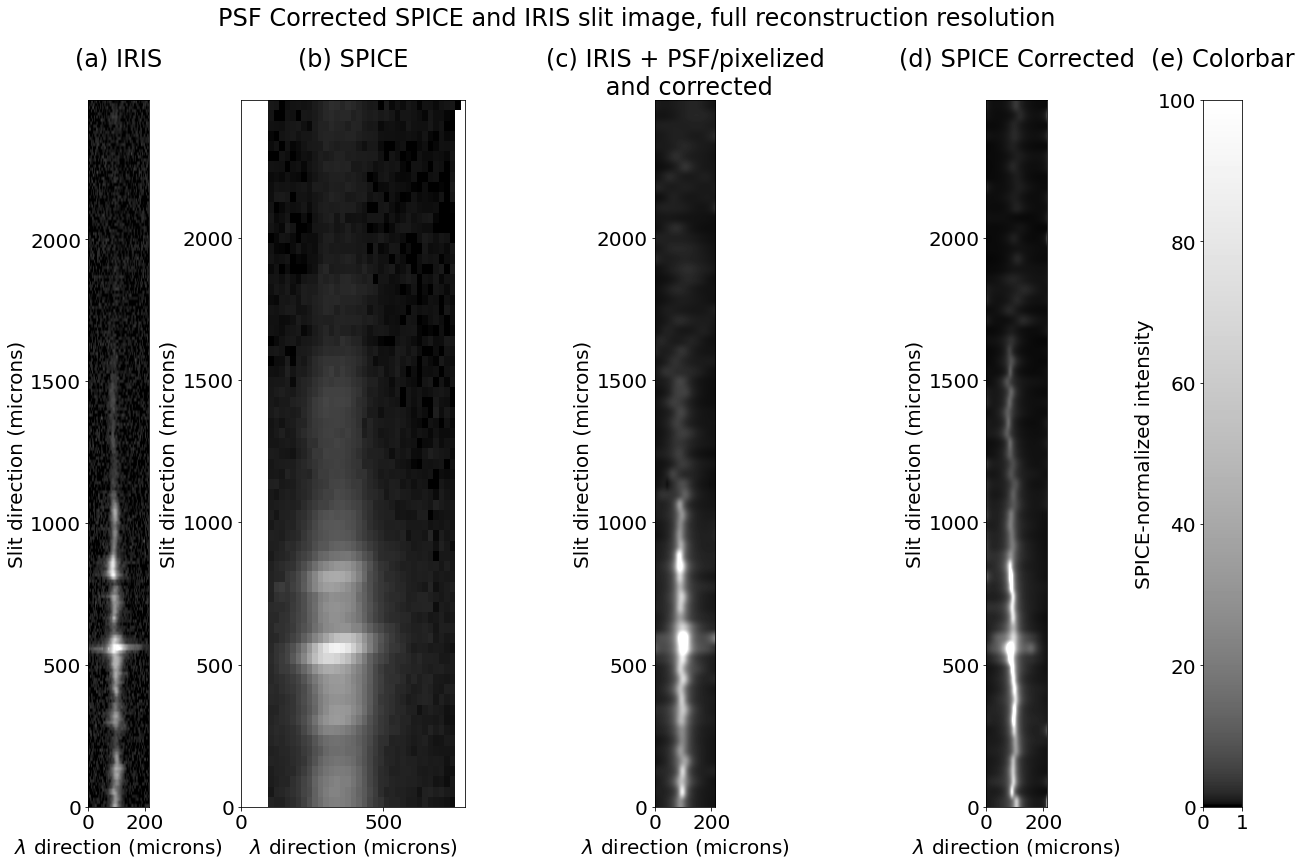}
			\end{center}
			\caption{High-resolution `raw' corrected spectra from the same slit position as is shown in \ref{fig:IRIS_SPICE_spectral_scan}. The IRIS data is at left (a), as before, while the SPICE data is now second from left (b). The correction applied to the degraded IRIS data is third from left (c), while the correction applied to SPICE is fourth from left (d). These last two are at higher spectral resolution than the input SPICE data; the higher resolution being necessary to resolve the spectral line. This over-resolution results in some noise-induced ringing and other ripples in the SPICE data, which we resolve by returning the image to SPICE pixel scale plus a `nominal' PSF, shown in Figure \ref{fig:corrected_slit_image}. Otherwise, the original IRIS line width is recovered, while the elongation and tilt of the SPICE PSF is removed.}\label{fig:hires_corrected_slit_image}
		\end{figure*}

\begin{figure*}[!htbp]
			\begin{center}
				\includegraphics[width=\textwidth]{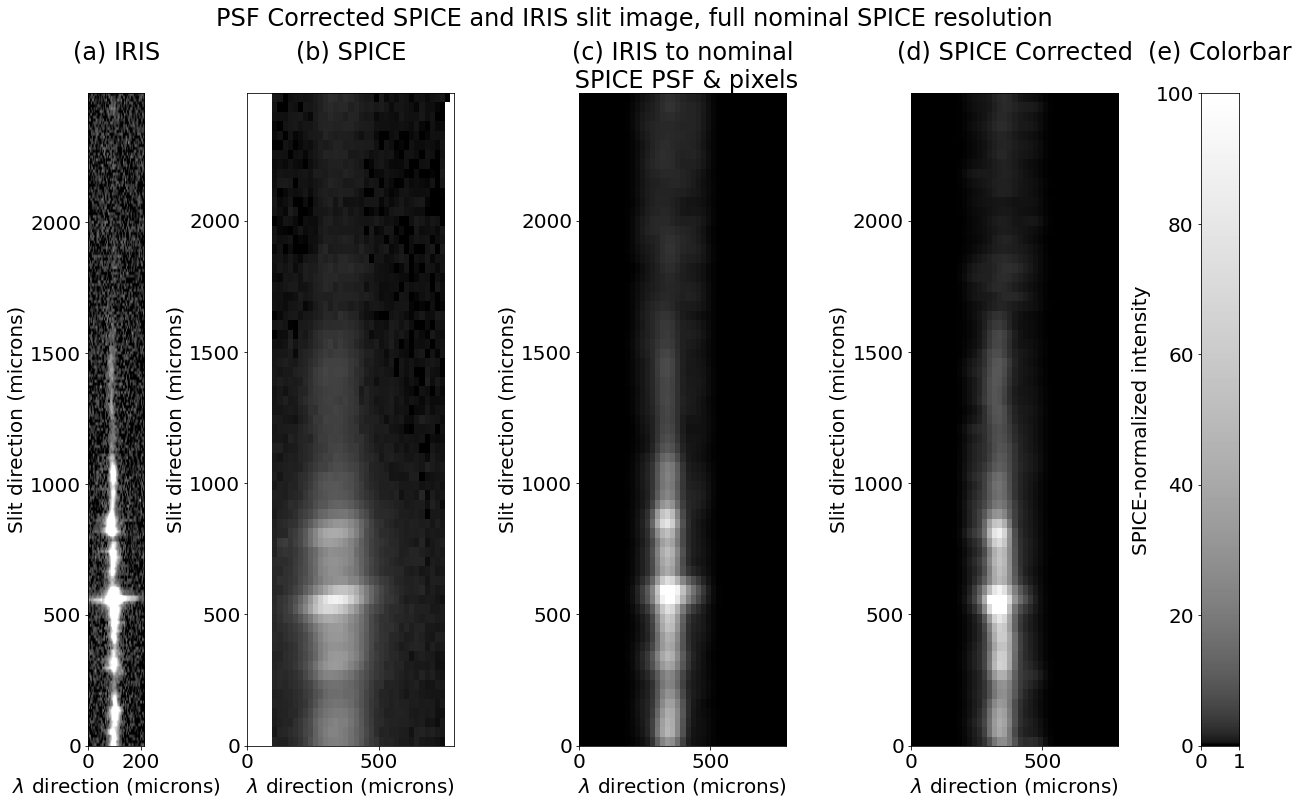}
			\end{center}
			\caption{Corrected spectra from the same slit position as is shown in \ref{fig:IRIS_SPICE_spectral_scan}. The IRIS data is at left (a), as before, while the SPICE data is now second from left (b). The correction applied to the degraded IRIS data is third from left (c), while the correction applied to SPICE is fourth from left (d). These last two have the same pixel scale as the input SPICE data, with a `nominal' PSF applied as well; while the `raw' corrected data (Figure \ref{fig:hires_corrected_slit_image}) is at higher resolution, which is necessary to resolve the spectral line, this over-resolution results in some noise-induced ringing and other ripples in the SPICE data. This is resolved in this figure by returning the image to SPICE pixel scale plus a `nominal' PSF. The original IRIS line width is recovered, while the elongation and tilt of the SPICE PSF is removed.}\label{fig:corrected_slit_image}
		\end{figure*}

		Figure \ref{fig:hires_corrected_slit_image} shows the same slit position as what's shown in Figure \ref{fig:IRIS_SPICE_spectral_scan}, with the higher spectral resolution, while Figure \ref{fig:corrected_slit_image} shows our final correction with this nominal/ideal SPICE PSF and pixelization. Each figure shows correction of both the SPICE data and the IRIS based `synthetic' SPICE data (IRIS with proposed SPICE PSF and pixelization). All show good removal of the rotation artifacts of the PSF, and the degraded then corrected IRIS data shows a good match to the originals (though not exact, which is to be expected considering the much lower resolution. The high res correction for SPICE (Figure \ref{fig:hires_corrected_slit_image}) shows the ringing and noise features mentioned, but these are absent when the resolution is reduced back the nominal SPICE values (Figure \ref{fig:corrected_slit_image}).

\begin{figure*}[!htbp]
			\begin{center}
				\includegraphics[width=\textwidth]{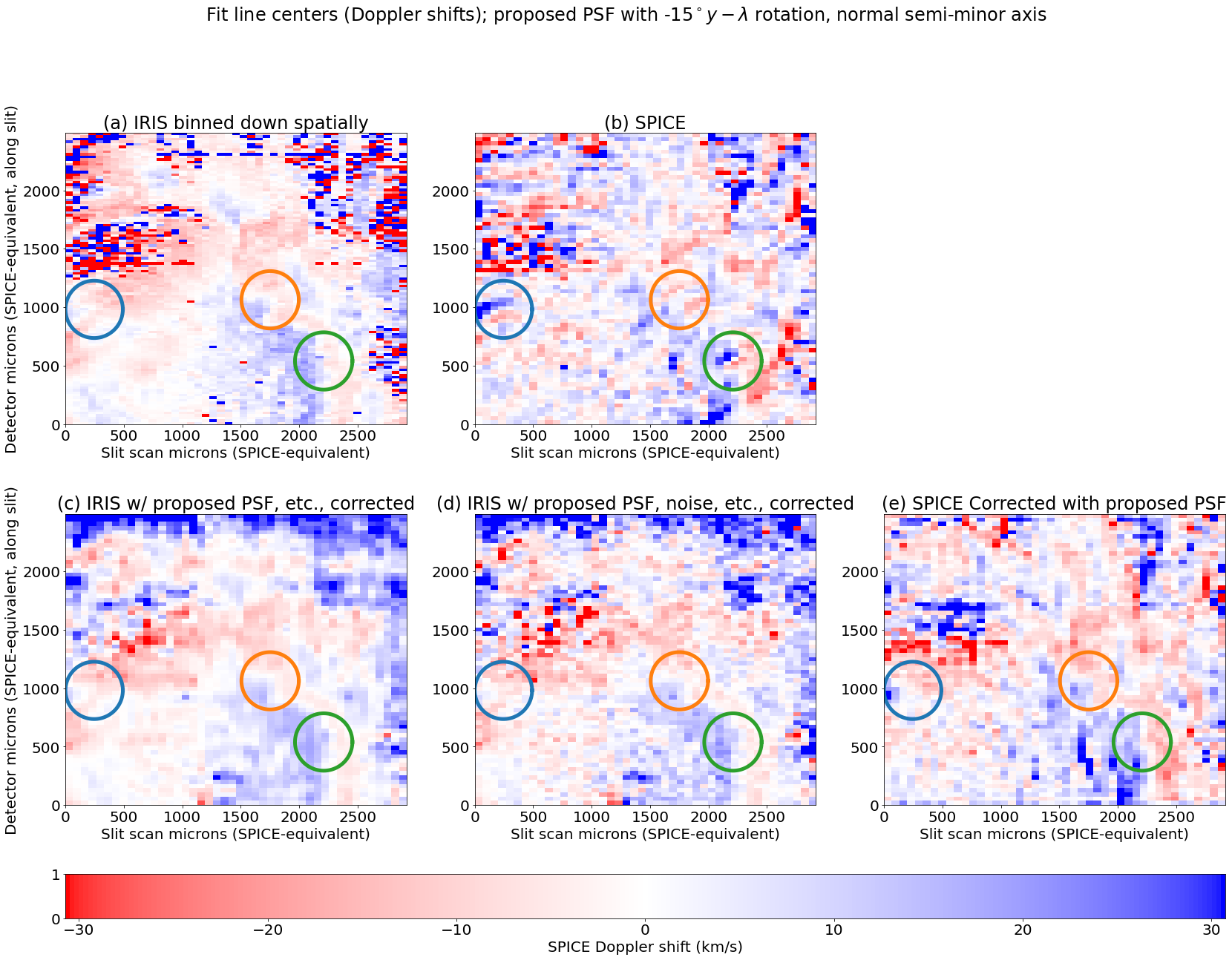}
			\end{center}
			\caption{Corrected Doppler shifts for both the degraded IRIS data and the SPICE data. Top left panel (a) shows the binned down IRIS data top center (b) shows original SPICE for comparison. The degraded, then corrected IRIS Doppler is shown at lower left (c), and is essentially identical to the original (with down-binning) IRIS data, which demonstrates that the correction is formally sound. The degraded, noise added, then corrected IRIS data is shown at lower center (d). Finally, the corrected SPICE data is shown at lower right (e), and looks very similar to the degraded with noise added IRIS data.  while the corrected SPICE Doppler looks very similar to the corrected IRIS and unlike the original SPICE data (for the degraded IRIS data, without added noise, and original SPICE data see Figure \ref{fig:example_doppler_fits}). This demonstrates that the correction also works on the real SPICE data.}\label{fig:doppler_artifacts_corrected}
		\end{figure*}

		More importantly, the PSF correction does remove the Doppler artifacts, both in the degraded then corrected IRIS data and in the SPICE data. This is shown in Figure \ref{fig:doppler_artifacts_corrected}. This figure has the degraded then corrected IRIS Doppler on left, the degraded IRIS Doppler in center, and the corrected SPICE on right (for original IRIS and SPICE of this data, see Figure \ref{fig:example_doppler_fits}). The degraded then corrected IRIS is essentially identical to the original (plus binned down) IRIS data, and the SPICE is quite similar to the IRIS. The artifacts previously seen at the circled locations are no longer present. There is some extra coarseness in the SPICE data, due to noise. There are also some artifacts in low signal regions which are likely due to a combination of noise and the interpolation applied to the SPICE L2 data, which we have been using. The interpolation can be mitigated by use of a an intermediate data product (with dark correction and flat fielding but no interpolation), but the low noise limit can't be -- it results in a loss of information that can't be recovered.

		\subsubsection{Correction applied to second region}\label{sec:correction_2ndregion}
		
\begin{figure*}[!htbp]
			\begin{center}
				\includegraphics[width=\textwidth]{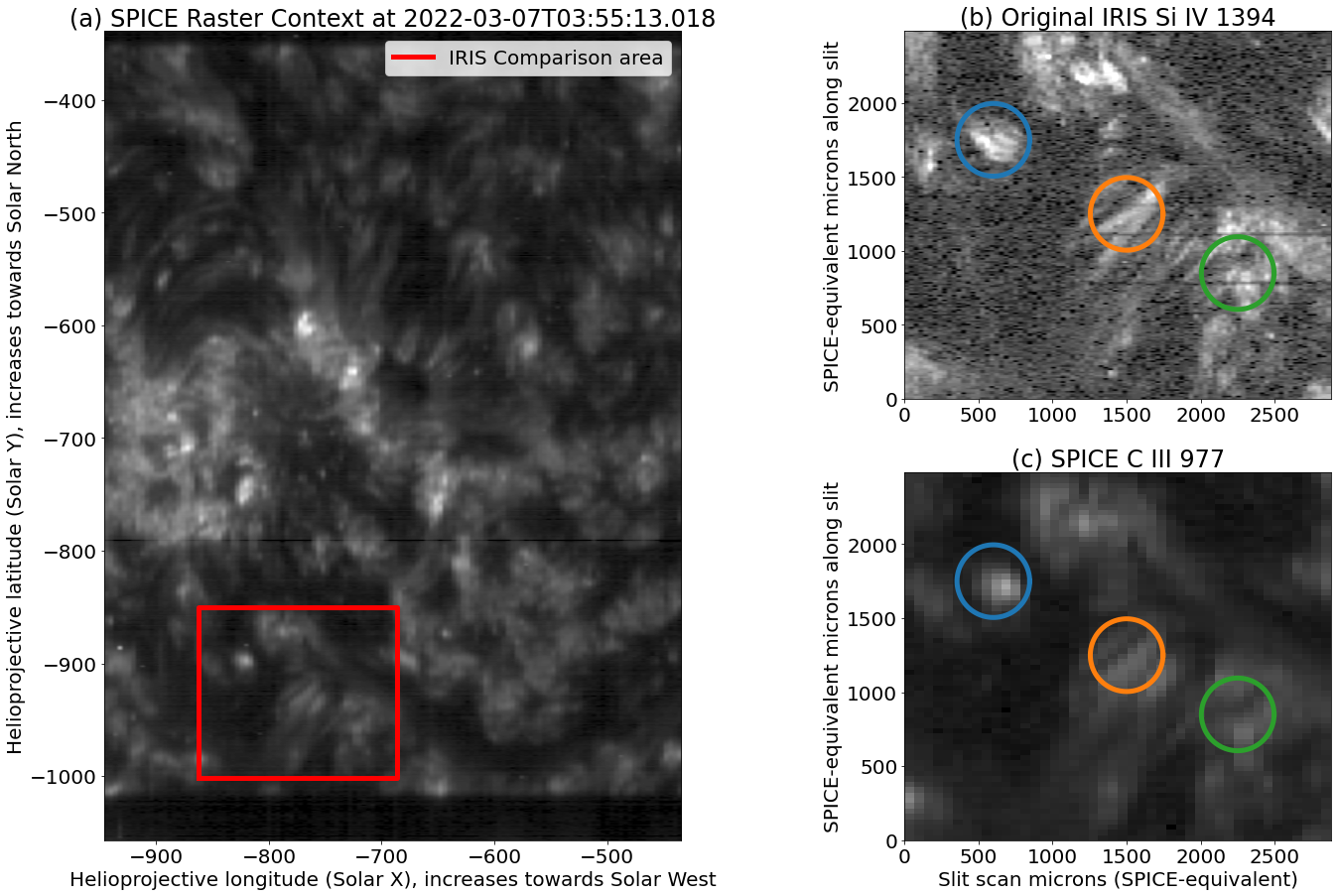}
			\end{center}
			\caption{Context image (a) of the SPICE-IRIS coordinated observation data for the second region considered. This is made from one of the full-size SPICE rasters; It is a spectral sum over the C III 977 \AA\ window. Detailed images of the line fit amplitudes in the region are shown at right (IRIS at top right, b, SPICE at lower right, c). Doppler fits are shown in Figure \ref{fig:corrected_spice_doppler_region2}.}\label{fig:context_image_region2}
		 \end{figure*}

\begin{figure*}[!htbp]
			\begin{center}
				\includegraphics[width=\textwidth]{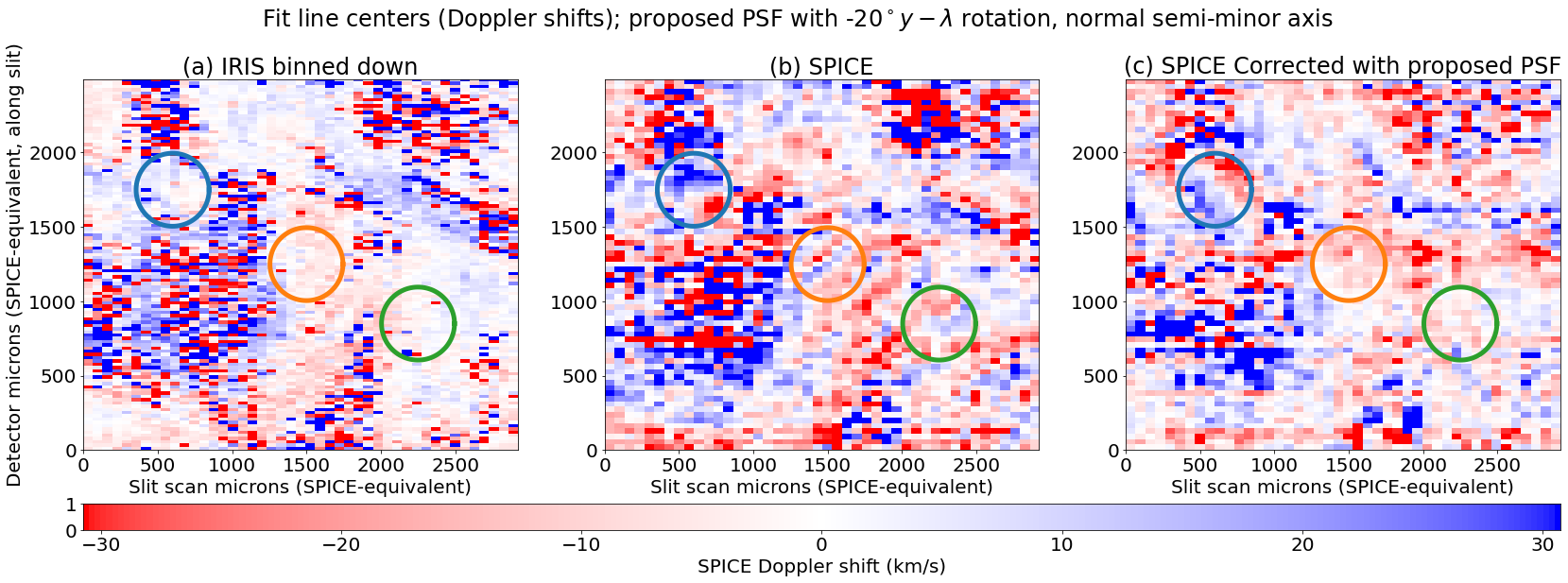}
			\end{center}
			\caption{Doppler shifts for a second region (see Figure \ref{fig:context_image_region2} for context) of the coordinated SPICE and IRIS observations. IRIS is shown at left (a). SPICE is shown at center (b), corrected SPICE is shown at right (c). This region has lower signal than the previous one, and a small highly complex region (blue circle at top left of each image) that shows a strong, likely artifact, feature in SPICE but not in IRIS. Generally the correction removes the artifacts although the agreement with IRIS is less strong than in the previous region (Figure \ref{fig:doppler_artifacts_corrected}). See text for further discussion.}\label{fig:corrected_spice_doppler_region2}
		\end{figure*}

		Last, we show correction of a second part of the coordinated IRIS and SPICE observing campaign. This region is more quiet sun, with a couple of isolated bright features (The mean and brightest features are both about a factor of 4 dimmer in this region than in the first one). Removal of artifacts from these features is another good test of the correction of the PSF. A context image for these observations is shown in Figure \ref{fig:context_image_region2}, while Doppler images for the region are shown in Figure \ref{fig:corrected_spice_doppler_region2}. The correction removes almost all the artifacts and results in a Doppler image similar to the IRIS data, although somewhat less so than before. A small relatively bright region (circled in blue) in the upper left of the region appears to still have some strong Doppler signals that are not present in IRIS, although they are reduced in magnitude. However, this region has a highly complex spectral structure and varies rapidly spatially and temporally in IRIS, so a close match between SPICE and IRIS isn't expected here. The corrected `synthetic' SPICE (IRIS degraded with noise added) do not show this artifact, so it would not seem to be an issue with the correction method itself. Nevertheless, it appears as if the `corrected' SPICE data may not be fully corrected in that location. Other differences can be attributed to the lower signal in the region and possibly also to more dissimilarities between the region as seen by IRIS vs. SPICE (the intensities appear more different than in the previous region). We have also had to increase the rotation angle of the PSF from $15^\circ$ to $20^\circ$ to best remove the artifacts; the rotation angle does appear to change across the field (we tried a variety of PSF rotation angles for both region, and the ones which best reduced the Doppler artifacts are shown). The correction can account for per-pixel and/or field-wide variation in the PSF, however; it is just a (perhaps not so simple) matter of quantifying that variation.

	\section{Conclusions}

		We have shown the presence of Doppler artifacts in the SPICE data and demonstrated that their likely cause is an elongated and rotated effective (i.e., $y-\lambda$) PSF. We have further demonstrated a novel method for correcting these artifacts, based on representing the forward transformation as a sparse matrix and then performing a regularized inversion of this matrix with a method based on $\chi^2$ minimization with $L^2$ norm and positivity constraint.
		
		The correction methods so described appears to work well. It reproduces the IRIS doppler shifts very well when applied to `synthetic' data (IRIS data with SPICE-like PSF and pixelization applied), and quantitatively in most places when applied to real SPICE data. There are some regions where the SNR is too low to correct the PSF and the reconstruction fails for that reason. There are fundamental information limitations to the ability to recover a high-resolution source from a blurred signal, but they do not appear to preclude correction of the Doppler artifacts and restoration of SPICE's nominal effective PSF. Other issues which presently affect the reconstruction include interpolation of the L2 data and uncertainty in the SPICE PSF, which we hope to rectify using modified data products and further coordinated observations with IRIS and Hinode EIS.
		
		It is also interesting in its applicability to a variety of other problems. The positivity constraint and regularization affords a degree of super-resolution capability, especially when the data being inverted are high-contrast, and so may also be useful for improving the sharpness of high-resolution imagery; unlike many sharpening methods, this method returns a true improved resolution version of its source, maintaining photometry, because its inversion remains consistent with the forward problem and original data. It should also be capable of inverting data from multiple instruments simultaneously, even those with differing resolutions. We may investigate reapplying the method to the DEM problem, for example for doing DEMs using data from AIA and Hinode XRT \citep{HinodeXRT} simultaneously. \cite{Plowman2021} applies a similar method to 3D reconstruction of the corona, demonstrating its flexibility.

		The corrections shown here were all done on a laptop computer with a 2.6 GHz 6 core Intel Core i7. Although the results shown have been reduced size, we have also carried out correction of full resolution data, on the same laptop. These took $\sim 2$ hours, considerably less than the average acquisition rate of SPICE data set as a whole. Therefore, the CPU load is easily within the capacity of a central server to reprocess the data, or of individual users to reprocess it on their own. Currently, the codebase has been provided to other members of the SPICE team for beta testing. Once this is finished, the code will be made publicly available. In the meantime, will be provided on request, pending approval of the SPICE team. The code are all written in Python, making use of Numpy \citep{numpy}, Astropy \citep{astropy, astropyII}, SciPy \citep{scipy}, and optionally SunPy \citep{sunpy}.
					
		\begin{acknowledgements}
		This work was supported by NASA under Goddard Space Flight Center subcontract \# 80GSFC20C0053 to Southwest Research Institute.
		\end{acknowledgements}
					
\bibliographystyle{aa}
\bibliography{psfrestore_paper0}

\end{document}